\documentclass[pra,twocolumn,showpacs,floatfix]{revtex4}
\usepackage{graphicx,amsmath,amsfonts,amssymb}
\usepackage[usenames]{color}

\begin{document}

\title{Localization of collisionally inhomogeneous condensates in a bichromatic
optical lattice}

\author{Yongshan Cheng$^{1,2}$\footnote{yong\_shan@163.com}
and
   S. K. Adhikari$^1$\footnote{adhikari@ift.unesp.br;
URL: www.ift.unesp.br/users/adhikari}}
\affiliation{$^1$Instituto de F\'{\i}sica Te\'orica, UNESP - Universidade
Estadual Paulista,
01.140-070 S\~ao Paulo, S\~ao Paulo, Brazil\\
$^2$Department of Physics, Hubei Normal University, Huangshi 435002,
People's Republic of  China
 }

\begin{abstract}
By direct numerical simulation and variational solution of the
 Gross-Pitaevskii equation, we studied the stationary
and dynamic characteristics of a cigar-shaped, localized,
collisionally inhomogeneous Bose-Einstein condensate trapped in a
one-dimensional bichromatic quasi-periodic optical-lattice
potential, as used in a recent experiment on the localization of a
Bose-Einstein condensate
 [Roati {\it et al.}, Nature (London) {\bf 453}, 895 (2008)]. The
effective potential characterizing the  spatially modulated
nonlinearity is obtained. It is found that the collisional
inhomogeneity has influence not only on the central region but also
on the tail of the Bose-Einstein condensate. The influence depends
on the sign and value of the spatially modulated nonlinearity
coefficient. We
also demonstrate the stability of the stationary localized state by
performing a standard linear stability analysis. Where possible, the
numerical results are shown to be in good agreement with 
the variational results.

\end{abstract}


\maketitle

\section{Introduction}
\label{I}

Since Anderson predicted a localization of noninteracting
electron wave in solids
with a disorder potential fifty  years ago
\cite{anderson}, the Anderson localization has been
observed and studied
extensively in optics \cite{optics}
and acoustics \cite{acous} and in Bose-Einstein condensates (BEC).
In the study of  Anderson localization in a BEC,
disorder  laser speckles \cite{billy} and
quasi-periodic optical lattices  (OL) \cite{roati}
have been used.
Random speckle
potentials are produced when light is reflected by a rough surface
or transmitted by a diffusive medium \cite{ chapter3}. 
Billy {\it et al.} \cite{billy} observed the exponential
tail of the spatial density distribution when a $^{87}$Rb BEC was
released into a one-dimensional (1D) waveguide in the presence of a
controlled disorder created by a weak laser speckle. A bichromatic OL is
realized by a primary lattice perturbed by a weak secondary lattice
with incommensurate wavelength \cite{PRL-98-130404}, and this
system corresponds to an experimental realization of the
Harper \cite{Harper} or Aubry-Andr\'e model \cite{Aubry}. Roati
\emph{et al}. \cite{roati} studied the localization of a noninteracting
$^{39}$K BEC in a bichromatic OL. There have been many theoretical
studies of
localization using the numerical solution of the
Gross-Pitaevskii (GP)  equation \cite{random}
as well as
using the Bose-Hubbard model \cite{roux} in addition to the
experimental studies under different conditions on disorder
\cite{chabe}.
There have been studies of localization in two- and three-dimensions
\cite{2D3D}
and of
 the destruction of localization with the
increase of nonlinear repulsion \cite{dnlse,adhikari1}.


A Feshbach resonance (FR)
driven by a magnetic \cite{nature-392-151} or optical
\cite{PRL-93-123001} field  allows one to vary   the atomic interaction
of a BEC in a controlled fashion
\cite{PRL-99-010403}, thus creating a noninteracting as well as
a weakly interacting BEC for the study of Anderson localization.
Fedichev {\it et al.} \cite{PRL-77-2913} predicted
that the spatial variation of the laser field intensity by proper choice
of the resonance detuning can lead to a spatial dependence of the
atomic interaction, creating the  so-called ``collisionally inhomogeneous"
BECs. The theoretical prediction has been demonstrated experimentally
by Theis  {\it et al.}
with the $^{87}$Rb BEC \cite{PRL-93-123001}.
Sakaguchi and Malomed studied the formation of solitons in such BECs
\cite{PRE-72-046610}.
There have been studies  of
matter-wave  bright and
dark
solitons of the cubic-quintic nonlinear Schr\"odinger
equation  with time- and
space-dependent  nonlinearities \cite{cubic-quintic}
in
a collisionally inhomogeneous environment,
and of
dynamical
effects of a bright soliton
BEC  with local and smooth space variations of the
two-body atomic scattering length  \cite{Abdullaev}.
There have been studies about how to introduce a space dependence
in the nonlinear interaction of the BEC in a controlled way
\cite{PRL-105-050405}. There have also been studies of soliton oscillations
\cite{existence}
and dynamical trapping and transport \cite{dynamics}
in collisionally inhomogeneous BECs.


Here we combine the two interesting
settings, namely, the bichromatic OL and a collisionally
inhomogeneous BEC (with a spatially modulated nonlinearity),
to study the statics and dynamics  of a localized BEC in this set up. We
assume that the spatial dependence of the nonlinearity,
{induced by the external magnetic field of the OL,}
has the same form
as the bichromatic OL, i.e., the nonlinear coefficient in the GP
equation is proportional to the OL
intensity \cite{PRE-72-046610, PLA-367-149}.
 We study the effects of the spatially modulated
nonlinearity on the shape of the density envelope,  and the stability
of the stationary localized states. The numerical
results are shown to be in good agreement with the
 variational results, where applicable. On the other hand, the tail
region of the stationary localized states is examined, where we
find exponential decay in space indicating localization in a weak
disorder potential.  We study the
location oscillation (oscillation of the center)
and breathing oscillation of the localized
states, induced by appropriate perturbation,
employing  numerical simulation of the GP equation.

In Sec. \ref{II} we present a brief account of the 1D GP equation and
the bichromatic OL potential used in our study, and a time-dependent
variational analysis of the GP equation under appropriate
conditions. We obtain a set of coupled evolution equations for the
parameters of the localized state. The effective potential characterizing
the spatially modulated nonlinearity is also described. In Sec.
\ref{III} we investigate the effects of the spatially modulated
nonlinearity on the central and tail regions of a stationary
localized BEC by a numerical solution of the GP equation
using
 the split-step Fourier spectral method. For
the central region, we demonstrate the stability of the localized
states by performing a standard linear stability analysis. In Sec.
\ref{IIII} the oscillation
dynamics of a collisionally inhomogeneous BEC
in a bichromatic OL is studied  numerically.
 In Sec. \ref{IIIII} we present a
brief discussion and concluding remarks.

\section{Analytical consideration of localization}

\label{II}

We consider a cigar-shaped quasi 1D
BEC with inhomogeneous atomic interaction
described by
wave function $u(x, t)$ satisfying  the
following dimensionless 
1D GP equation \cite{kivshar,1d}
\begin{eqnarray}\label{eq1}
i\frac{\partial u}{\partial t}=- \frac{1}{2}\frac{\partial^2
u}{\partial x^2} +g(x)|u|^2u+V(x)u,
\end{eqnarray}
with normalization $\int_{-\infty}^{\infty}|u|^2 dx=1$. 
The space variable
$x$, time $t$, and energy are expressed in transverse harmonic
oscillator units $a_\perp=\sqrt{\hbar/(m\omega)}$, $\omega^{-1}$ and
$\hbar\omega$, where $m$ is the mass of an atom and $\omega$ is the
angular frequency of the transverse trap. As in the experiment of
Roati {\it et al.} \cite{roati}, the potential $V(x)$ is taken to be  a
bichromatic OL
of incommensurate wave lengths:
\begin{eqnarray}\label{pot}
V(x)=\sum_{l=1}^2 A_l \sin^2(k_lx),
\end{eqnarray}
with $A_l=2\pi^2 s_l/\lambda_l^2, (l=1,2)$, where $\lambda_l$'s are
the wavelengths of the OL potentials, $s_l$'s are their intensities,
and $k_l=2\pi/\lambda_l$ the corresponding wave numbers.
In this
investigation, we take the incommensurate ratio of the two
components \cite{modugno} $k_2/k_1=(\sqrt{5}-1)/2\approx 0.618033989...$.
In the actual experiment of Roati {\it et al.}
\cite{roati}, however, the parameter is $k_2/k_1=1.1972...$. Without
losing generality, we further take $\lambda_1=10$, and $s_1=10$,
$s_2=0.3s_1$ which is roughly the same ratio $s_2/s_1$  as in the experiment
of Roati {\it et al.} \cite{roati}.

By means of the FR technique controlled by properly designed
configurations of external optical fields
\cite{PRE-72-046610, PLA-367-149}, the spatial variation of laser
field intensity $I(x)$ produces the spatial variation of the atomic
scattering length. As the potential $V(x)$ is also proportional to
laser field intensity, it is reasonable to assume the
spatial-dependence of the atomic scattering length is similar to
$V(x)$.  Then, the atomic scattering length can be given as
$a_s=a_{s0}+cV(x)$; here $V(x)$ is the same as Eq. (\ref{pot}),
$a_{s0}$ is scattering length of the corresponding collisionally
homogeneous system, and $c$ is a constant coefficient related
to the optical intensity and may be either positive or negative.
Thus, the nonlinear coefficient $g(x)$ in Eq. (\ref{eq1}) has a
spatial dependence of the form,
\begin{eqnarray}\label{g}
g(x)=\varepsilon_0+\varepsilon\sum_{l=1}^2 A_l \sin^2(k_lx).
\end{eqnarray}
The nonlinearity $\varepsilon_0$ is given by \cite{1d}
$\varepsilon_0=2a_{s0}N/a_\perp$ with $N$ the number of atoms, 
and {
$\varepsilon=2cN/a_\perp$}
is the spatially modulated nonlinearity coefficient.  Because the
intensity-independent nonlinear coefficient $\varepsilon_0$ may be
altered independently of the other parameters \cite{PRL-99-010403},
in order to focus our attention on the effect of the spatially
modulated nonlinearity on the localization of the BEC, we let
$\varepsilon_0=0$ in the following.

Usually, the BEC localized states formed on a  bichromatic lattice
may occupy many sites of OL potential  \cite{pre-47-1375,roati}. For
certain values of the parameters, however, potential (\ref{pot})
leads to localized states confined practically to a single site of
the OL potential. When this happens, a variational approach for
solving the GP equation is useful.
 To apply the variational approach  to the localized BEC,
  we adopt the following
variational ansatz
\begin{eqnarray}\label{va}
u(x,t)&=&\frac{1}{\pi^{1/4}}\sqrt{\frac{N}{w}}
\exp\left[-\frac{(x-x_0)^2}{2w^2}\right] \nonumber \\
&&\times\exp\left\{i\left[\gamma(x-x_0)+\beta(x-x_0)^2+\phi\right]\right\}
,
\end{eqnarray}
with $w$  the spatial width of the localized state centered at
$x_0$, $\gamma$  the linear phase coefficient, $\beta$  the chirp
 and $\phi$  the phase. These are
time-dependent parameters \cite{pre-47-1375}. The Lagrangian of the system
is \cite{PG,cheng}
\begin{eqnarray}
&L(t)&=\int_{-\infty}^{\infty}\Biggl[\frac{i}{2}\left(u^\ast\frac{\partial{u}}{\partial{t}}
-u\frac{\partial{u}^\ast}{\partial{t}}\right)
-\frac{1}{2}\left|\frac{\partial{u}}{\partial{x}}\right|^2 \nonumber \\
&&-\frac{1}{2}g(x)\left|u\right|^4-V(x)\left|u\right|^2\Biggr]dx
  \nonumber \\
 &=&N(\gamma\dot{x}_0-\frac{1}{2}\dot{\beta}w^2-\dot{\phi})-\frac{N}{2}
 \left(\frac{1}{2w^2}+\gamma^2+2\beta^2w^2\right)\nonumber \\
 &&+N^2L_M+NL_V,
\end{eqnarray}
where the overhead
dot denotes time derivative, the star denotes
complex conjugation and
\begin{eqnarray}
L_M&=&\frac{\varepsilon}{4\sqrt{2\pi}w}\sum_{l=1}^2A_l\left[
\cos(2k_lx_0)\exp\left(-\frac{k_l^2w^2}{2}\right)-1\right],\nonumber \\
 \\
L_V&=&-\frac{1}{2}\sum_{l=1}^2A_l\left[1-
\cos(2k_lx_0)\exp\left(-k_l^2w^2\right)\right].
\end{eqnarray}
We use the variational Euler-Lagrange equation
\begin{eqnarray}
\frac{\partial L}{\partial \sigma}-\frac{d}{dt}\frac{\partial
L}{\partial\dot{\sigma}}=0,
\end{eqnarray}
where the variational parameters are $\sigma=\phi, x_0, \gamma,
\beta, $ and $w$. The first
variational equation using $\sigma=\phi$ yields $N=$constant.
We  take this constant to be unity
and use it in the subsequent equations. The other choices of $\sigma$
respectively, lead to  the following equations
\begin{eqnarray}
&\dot{\gamma}&= -\frac{\varepsilon}{2\sqrt{2\pi}w}\sum_{l=1}^2A_lk_l
\sin(2k_lx_0)\exp\left(-\frac{1}{2}k_l^2w^2\right)\nonumber \\
&&-\sum_{l=1}^2A_lk_l\sin(2k_lx_0)\exp\left(-k_l^2w^2\right),\label{velo} \\
&\dot{x}_0&= \gamma,\label{posi}\\
&\dot{w}&= 2\beta w\equiv F(w,\beta), \label{width} \\
&\dot{\beta}&= \frac{1}{2w^4}-2\beta^2+\frac{1}{w}\frac{\partial
L_M}{\partial w}+\frac{1}{w}\frac{\partial L_V}{\partial w}\equiv
G(w,\beta).\label{chirp}
\end{eqnarray}

The Hamiltonian of the BEC  is
\begin{eqnarray}\label{hami}
H&=&\dot{x}_0\frac{\partial L}{\partial\dot{x}_0}+
\dot{\phi}\frac{\partial L}{\partial \dot{\phi}}+
\dot{\beta}\frac{\partial L}{\partial \dot{\beta}}-L \nonumber \\
&=&\frac{1}{2}\left(\frac{1}{2w^2}+\gamma^2+2\beta^2w^2\right)-L_M-L_V.
\end{eqnarray}
Equations (\ref{width}) and (\ref{chirp}) determine the evolution
of the width $w$
once $x_0$  is known.
Equations (\ref{velo}) and (\ref{posi}) determine the evolution of
center $x_0$ once the width $w$ is known.
To study the dynamics of the localized state,
we insert Eq. (\ref{velo}) into Eq.
(\ref{posi}) to get an  anharmonic effective potential $V_{\rm{eff}}$:
\begin{eqnarray}\label{veff}
\frac{d^2x_0}{dt^2}&=&-\frac{\partial V_{\rm{eff}}}{\partial
x_0}\equiv-\frac{\partial}{\partial
x_0}\left(V_{\rm{effM}}+V_{\rm{effV}}\right),\\
\label{veffM}
V_{\rm{effM}}&=&\frac{-\varepsilon}{4\sqrt{2\pi}w}\sum_{l=1}^2A_l
\cos(2k_lx_0)\exp\left(-\frac{k_l^2w^2}{2}\right),\nonumber \\ \\
V_{\rm{effV}}&=&-\frac{1}{2}\sum_{l=1}^2A_l\cos(2k_lx_0)\exp\left(-k_l^2w^2\right).
\label{veffV}
\end{eqnarray}
The effective potential has two terms. The second term on the
right of Eq. (\ref{veff}), $V_{\rm{effV}}$, arises from the
bichromatic OL and contributes to an attractive
well, if $|x_0|$ is small enough. The first term, $V_{\rm{effM}}$,
is induced by the spatial modulation of the nonlinearity and is
actually a pseudo-potential \cite{PRA-78-053601}.
The pseudo potential is quasi
periodic  and is a potential barrier or well depending on
the sign and value of the coefficient $\varepsilon$.


\section{Stationary Localized State}
\label{III}

\begin{figure}
\begin{center}
\includegraphics[width=\linewidth]{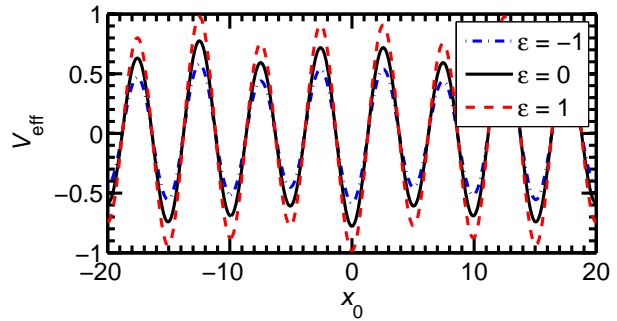}
\end{center}

\caption{(Color online) The dimensionless effective potentials $V_{\rm eff}$
``felt" by an immobile localized state for different $\varepsilon$
from Eqs. (\ref{veffM}) and (\ref{veffV}) where the width $w$ is 
calculated from Eq. (\ref{width0}). } \label{fig1}
\end{figure}

The stationary states are obtained
by setting the time derivative in Eqs. (\ref{velo}) $-$ (\ref{chirp})
to zero. Then the simplest solution of
 Eqs. (\ref{velo}) and (\ref{posi})
is $x_0=0$ and  we consider below
 an immobile localized state at
origin ($x_0=0$) with $\gamma=0$. Equations (\ref{width}) and
(\ref{chirp})  determine $w$ and can be written as
\begin{eqnarray}
\beta_0&=&0, \label{chirp0}\\
1&+&\frac{\varepsilon
w_0}{2\sqrt{2\pi}}\sum_{l=1}^2A_l\left[1-\left(1+k_l^2w_0^2\right)
\exp\left(-\frac{1}{2}k_l^2w_0^2\right)\right]\nonumber \\
&-&2w_0^4\sum_{l=1}^2A_lk_l^2\exp\left(-k_l^2w_0^2\right)=0.\label{width0}
\end{eqnarray}
 The effective potential felt by a
stationary localized state at $x=x_0=0$ is obtained from Eqs.
(\ref{veff}) $-$ (\ref{veffV}) and is
plotted in Fig. \ref{fig1} where the
width $w$ is obtained by solving numerically Eq. (\ref{width0}).
With the increase of $\varepsilon$, the
strength of  both disorder and the quasi
periodic effective potential increases.


To understand the effects of the coefficient $\varepsilon$, we
obtain the stationary localized states by solving Eq.
(\ref{eq1}) numerically  with real-time Split-step Fourier
spectral method
with a space step  0.04 and time step
0.0001.  We
checked the accuracy of the results by varying the space and time
steps and the total number of space and time steps.
To compare with numerics,
 the variational width is obtained by solving Eq.
(\ref{width0}).

\begin{figure}
\begin{center}
\includegraphics[width=.49\linewidth]{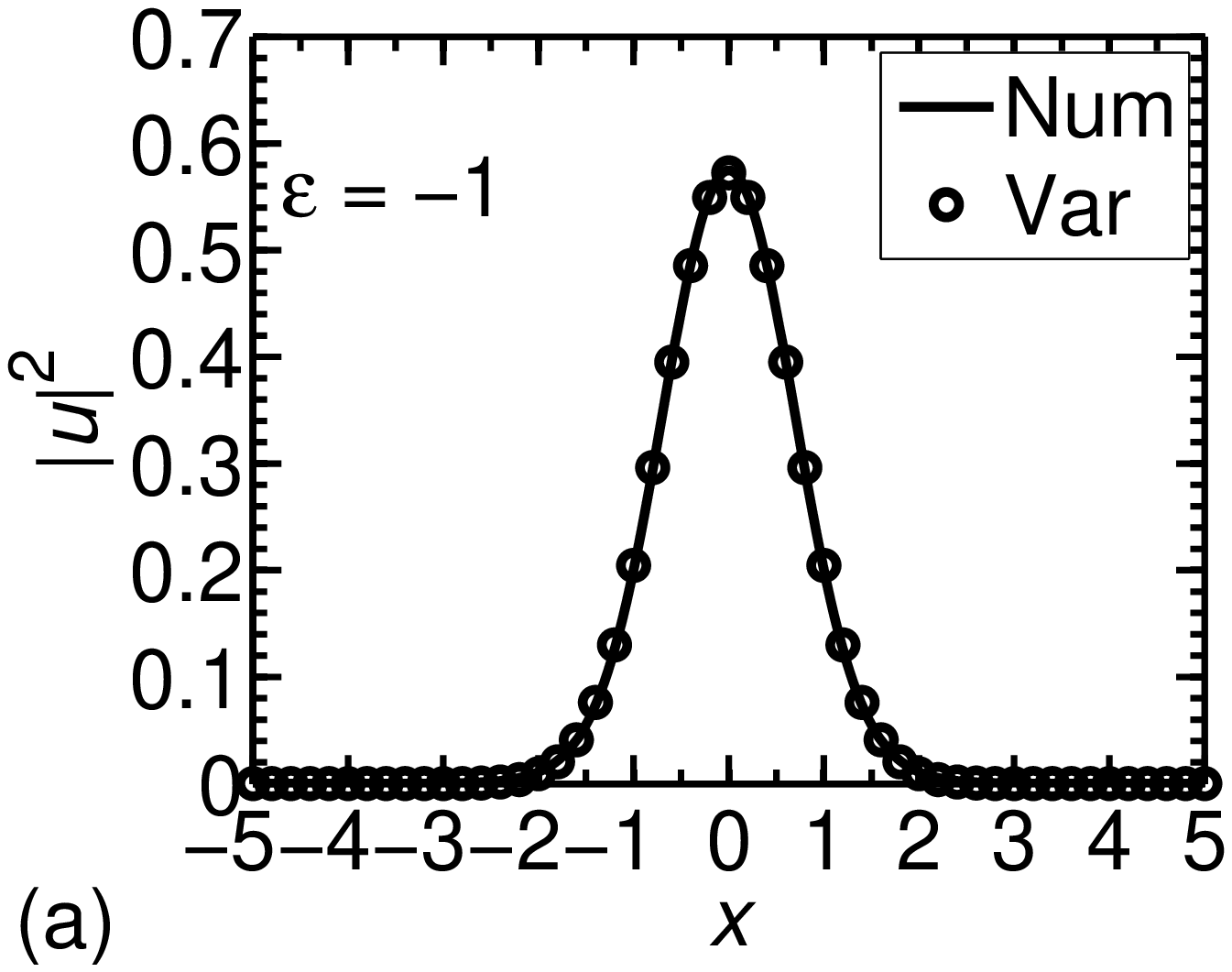}
\includegraphics[width=.49\linewidth]{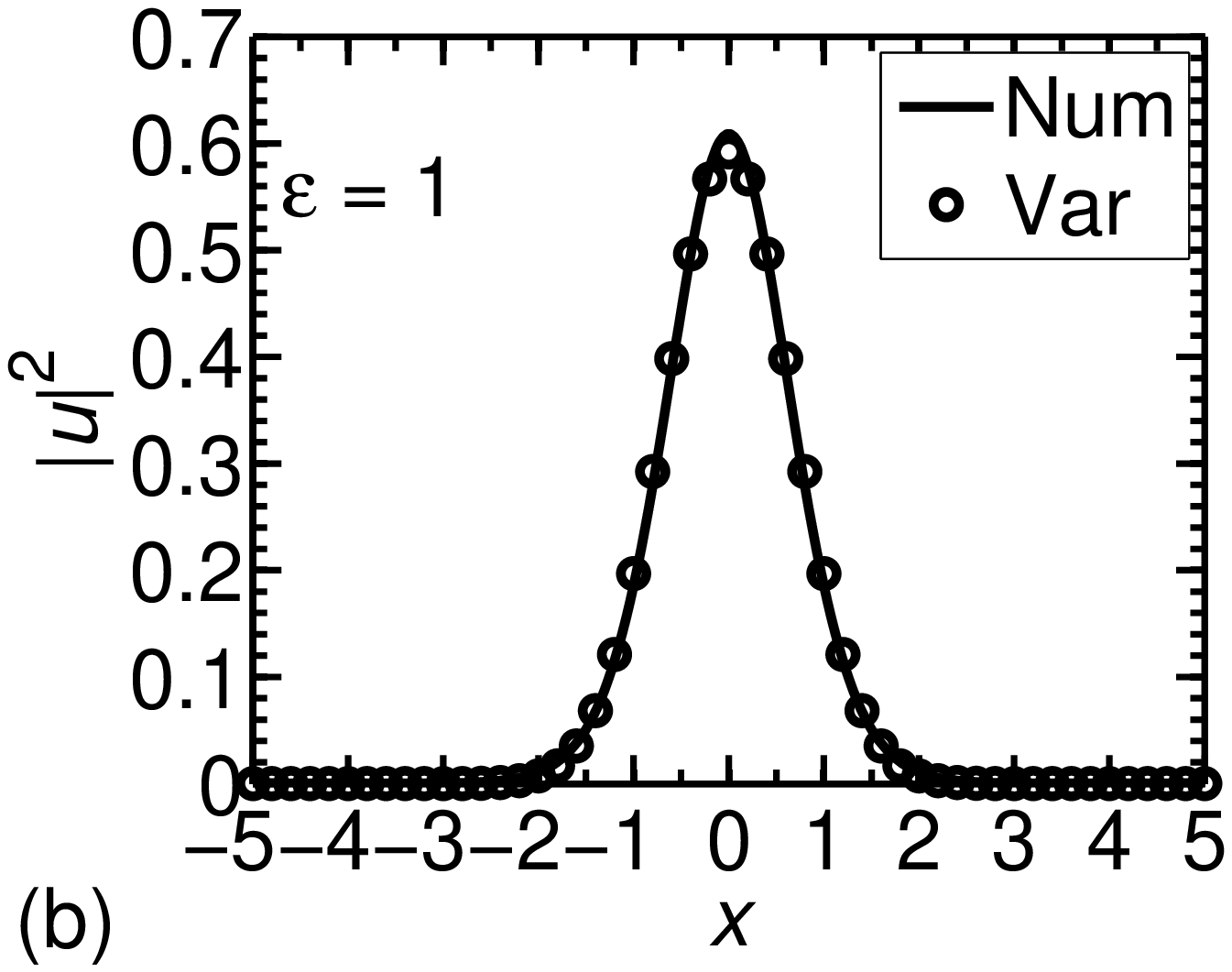}
\includegraphics[width=.49\linewidth]{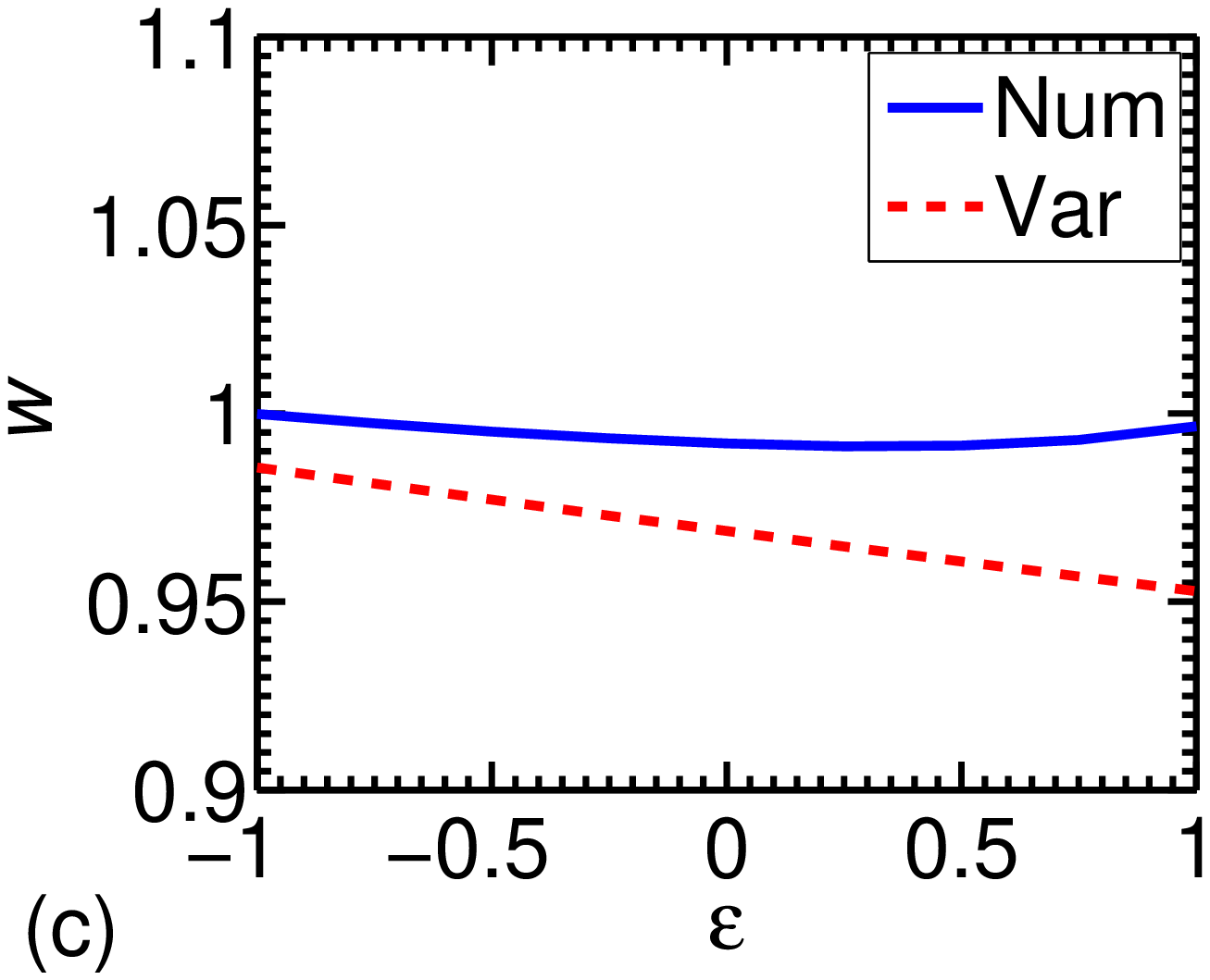}
\includegraphics[width=.49\linewidth]{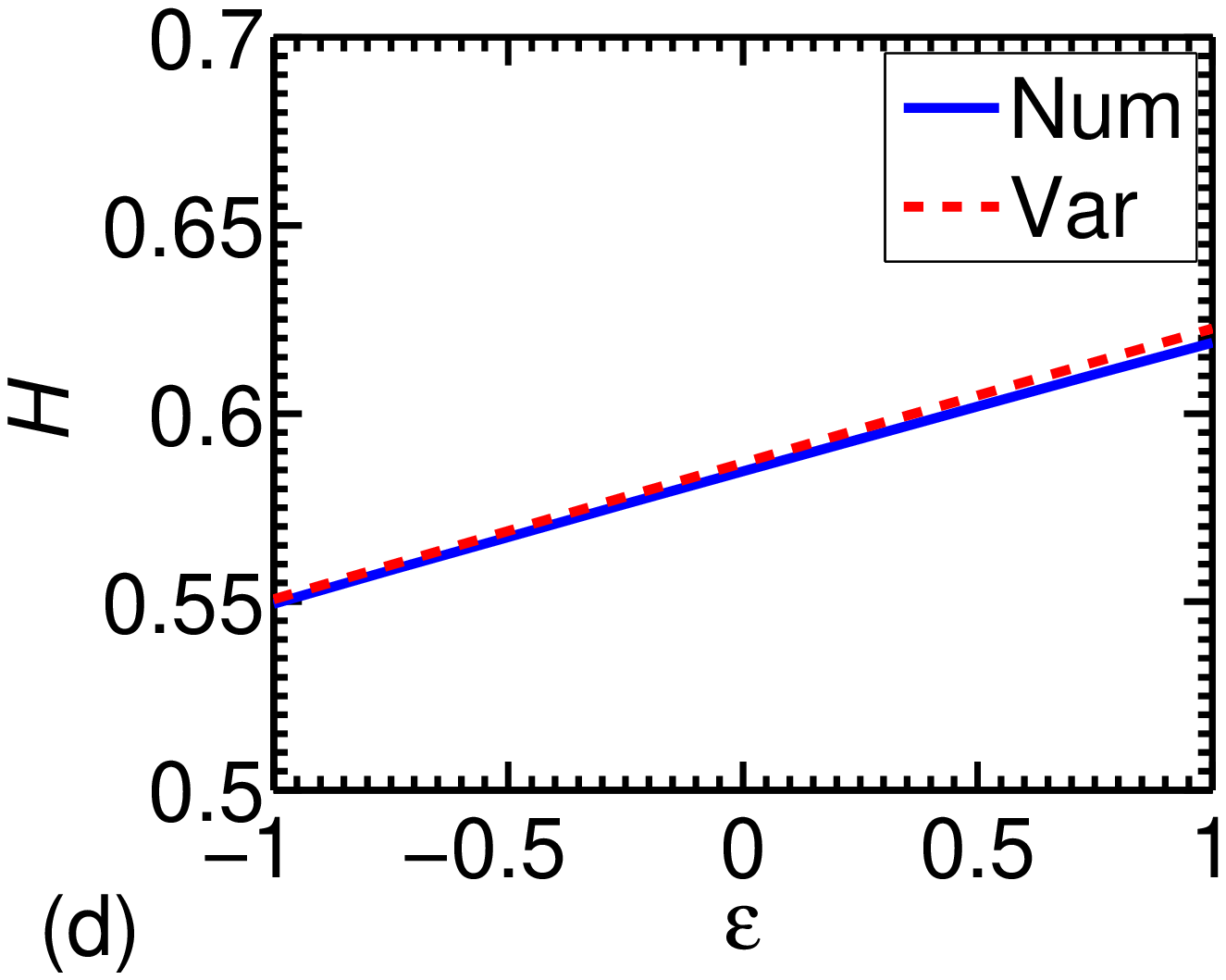}
\end{center}

\caption{(Color online) Numerical (lines) and variational (chain of
symbols) 
densities $|u|^2$ of the localized BEC versus $x$ for (a)
$\varepsilon=-1,$ (b)  $\varepsilon=1$. (c)  Numerical (solid lines)
and variational (dashed lines) dimensionless 
widths $w$ versus  $\varepsilon$ and
(d) Hamiltonian versus $\varepsilon$. (All quantities are dimensionless.)}
\label{fig2}
\end{figure}

Typical numerical and variational results for density, width and
Hamiltonian of the localized state at $x=0$
are exhibited in Fig. \ref{fig2}
for  $-1\leq\varepsilon\leq 1$.
The variational Hamiltonian is obtained from Eq. (\ref{hami}) and
the numerical Hamiltonian  is obtained from
 $H=
\int_{-\infty}^{\infty}[|u'|^2/2+g(x)|u|^4/2+V(x)|u|^2]dx$.
 Figures \ref{fig2} (a) and (b) exhibit the density of the stationary
localized states corresponding to $\varepsilon=-1$ and 1, respectively.
Figures \ref{fig2} (c) and (d) exhibit the variation of $w$ and $H$ with
$\varepsilon$. The numerical width $w$ in Fig. \ref{fig2} (c) is 
{{$\sqrt 2$ times}}
the 
root-mean-square (rms)
size of the BEC. Figures \ref{fig2} indicate that the
numerical results are in good agreement with the variational results.
Figure \ref{fig2} (c) shows that the width decreases and Fig. \ref{fig2}
(d) shows that the Hamiltonian increases with the change of
$\varepsilon$ from negative to positive. 
{{The dependence of the variational 
width on $\varepsilon$ can be 
qualitatively understood as follows. 
The
height of the central well of the 
effective potential increases with increasing
$\varepsilon$ as shown in Fig. \ref{fig1}. Hence the central part of the 
localized state with a Gaussian shape becomes narrow with the increase of  
$\varepsilon$, as can be seen from Figs. \ref{fig2} (a) and (b). 
The variational Gaussian ansatz only represents this central 
part and hence the variational width decreases with increasing $\varepsilon$.
However,  the numerical width (rms size)
shown in Fig. \ref{fig2} (c) receives nontrivial  
contributions from both the central Gausian part and the extended exponential 
tail  of the wave function (viz. Fig. \ref{fig4} 
(a)), making it difficult to predict even qualitatively the  
variation of the 
numerical width with $\varepsilon$.}}
It is interesting to compare these
results with those of a collisionally homogeneous condensate where a
constant negative (attractive) nonlinearity leads to a reduction of the
width and a constant positive (repulsive) nonlinearity increases the
width of the localized state \cite{adhikari1}. The numerical width is
larger than the variational width in Fig. \ref{fig2} (c) due to the long
exponential tail of the numerically obtained BEC (viz., Fig.
\ref{fig4}).  Also, the difference between the variational and numerical
widths increases for larger values of $\varepsilon$, because for larger
$\varepsilon$ the exponential tail is more pronounced
 resulting in a larger width
(viz., Fig.
\ref{fig4}).

\begin{figure}
\begin{center}
\includegraphics[width=\linewidth]{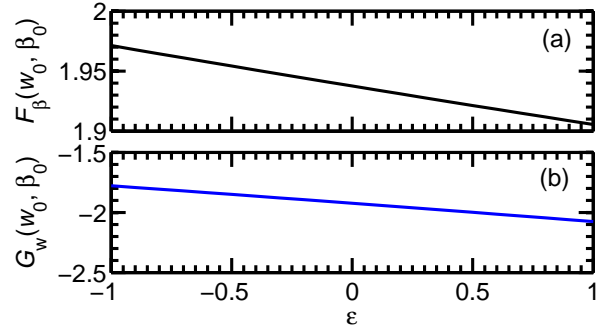}
\end{center}

\caption{(Color online) (a) The dimensionless 
functions $F_\beta(w_0, \beta_0)$
and (b) $G_w(w_0, \beta_0)$  versus
$\varepsilon$. It shows that  the signs of the two functions
are opposite for $-1 \leq \varepsilon \leq 1$.}
\label{fig3}
\end{figure}

It is important to investigate the stability of the stationary state
governed  by Eqs. (\ref{chirp0}) and (\ref{width0}) against
perturbation  by  a standard linear stability analysis.
Introducing small fluctuations around the stationary solution $(w_0,
\beta_0)$, $w'(t)=w(t)-w_0, \beta'(t)=\beta(t)-\beta_0$, and
linearizing Eqs. (\ref{width}) and (\ref{chirp}) indicated by them,
a set of two linear equations are obtained:
\begin{eqnarray}
\frac{d w'(t)}{dt}&=&F_w(w_0, \beta_0)w'(t)+F_\beta(w_0, \beta_0)\beta'(t),\\
\frac{d \beta'(t)}{dt}&=&G_w(w_0, \beta_0)w'(t)+G_\beta(w_0,
\beta_0)\beta'(t),
\end{eqnarray}
where the  subscripts $w$ and $\beta$  
denote a derivative with respect to the respective 
variable. Assuming the solution of $w'(t)$ and $\beta'(t)$ in
exponential form, $\sim \exp({\cal E} t)$, the eigen value ${\cal E}$
is
\begin{eqnarray}
2{\cal E}&=&F_w(w_0, \beta_0)+G_\beta(w_0, \beta_0) \nonumber \\ &&
\pm\Bigl\{\left[F_w(w_0, \beta_0)+G_\beta(w_0, \beta_0)\right]^2  \nonumber \\
&&+4F_\beta(w_0, \beta_0)G_w(w_0, \beta_0)\Bigr\}^{1/2}.
\end{eqnarray}
From Eqs.  (\ref{width}) and (\ref{chirp}) , we find
$F_w(w_0, \beta_0)=0, G_\beta(w_0, \beta_0)=0,$ and
\begin{eqnarray}
&& F_\beta(w_0, \beta_0)=2w_0, \label{Fbeta}\\
&& G_w(w_0, \beta_0)=-\frac{2}{w_0^5}+2w_0\sum_{l=1}^2A_lk_l^4\exp\left(-k_l^2w_0^2\right)\nonumber \\
&&\;\;\;\;
+\frac{\varepsilon}{4\sqrt{2\pi}w_0^4}\sum_{l=1}^2A_l\Biggl[-3
+\left(3+2k_l^2w_0^2+k_l^4w_0^4\right)\nonumber \\
&&\;\;\;\;\times\exp\left(-\frac{1}{2}k_l^2w_0^2\right)\Biggr],\label{Gw}
\end{eqnarray}
which leads to the eigen-values
\begin{eqnarray}\label{eigen}
  {\cal E}=\pm\left[F_\beta(w_0, \beta_0)G_w(w_0, \beta_0)\right]^{1/2}.
\end{eqnarray}
To investigate the stability, Eq. (\ref{width0}) is
first solved to get $w_0$ as a function of $\varepsilon$. This
result is then inserted in Eqs. (\ref{Fbeta}) and (\ref{Gw}) to get
$F_\beta(w_0, \beta_0)$ and $G_w(w_0, \beta_0)$. The graphical
representation of the two functions is shown in Figs. \ref{fig3} (a)
and (b). In the case of a small coefficient $\varepsilon$,
we can find that $G_w(w_0, \beta_0)<0$
 and $F_\beta(w_0, \beta_0)=2w_0>0$. Thus,
both the eigen-values from Eq. (\ref{eigen}) must be
imaginary, and the localized state from  Eqs. (\ref{chirp0}) and
(\ref{width0}) is stable against small perturbation.

\begin{figure}
\begin{center}
\includegraphics[width=.49\linewidth]{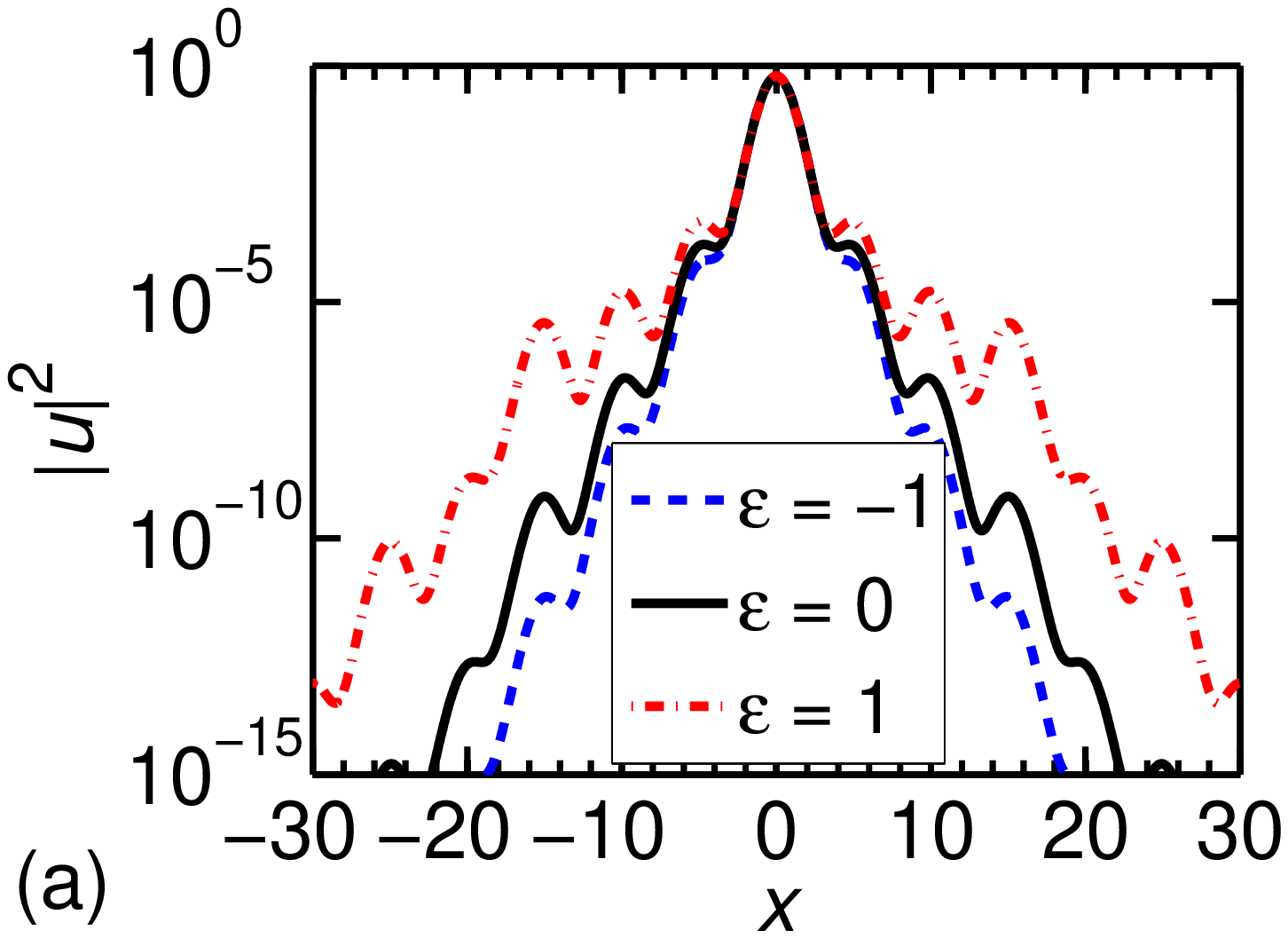}
\includegraphics[width=.49\linewidth]{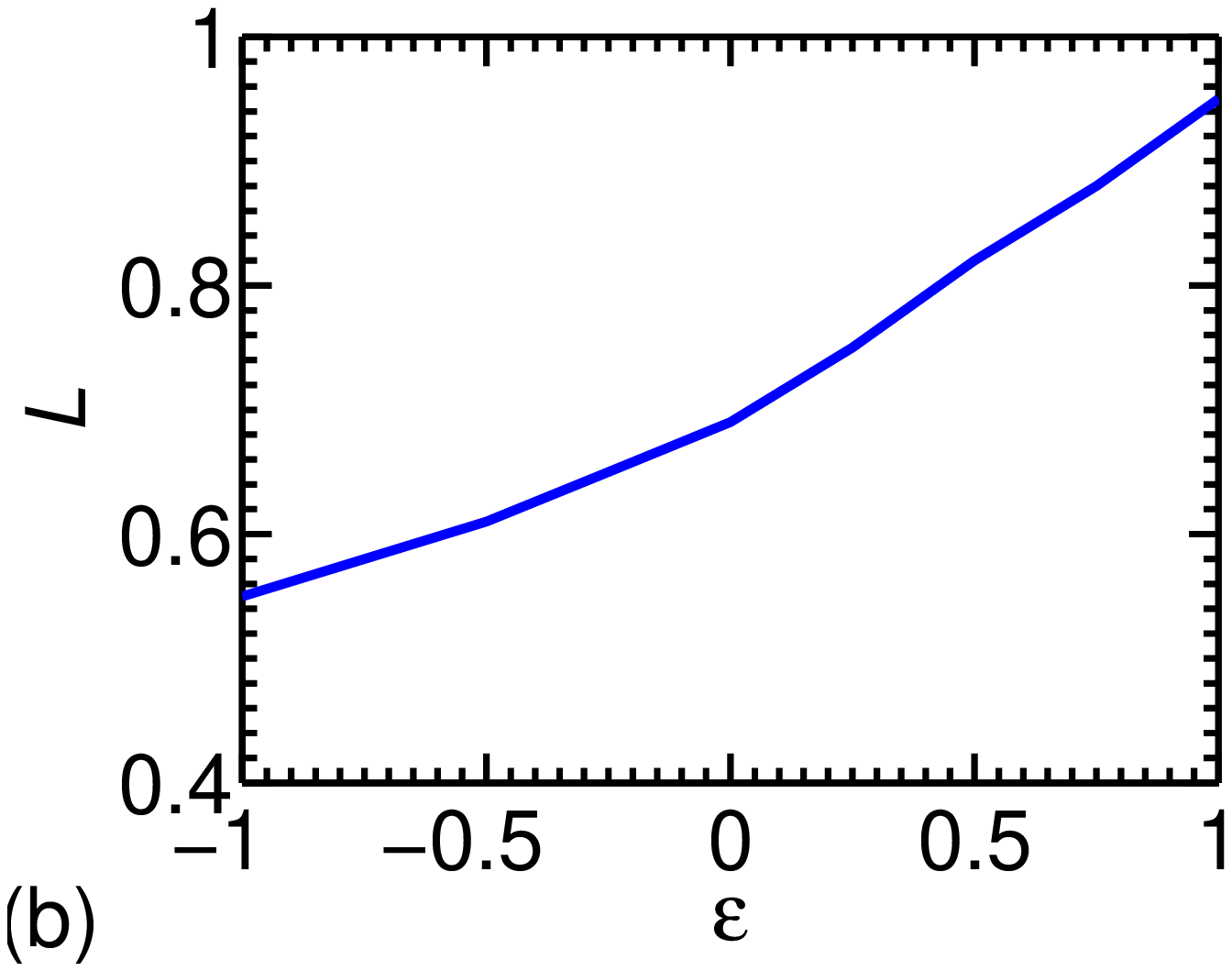}
\end{center}

\caption{(Color online) (a) The numerically obtained 
dimensionless density $|u|^2$ versus dimensionless 
$x$ for  different  $\varepsilon$.
(b) The dimensionless localization lengths $L$ versus
 $\varepsilon$. The localization length $L$ is obtained
by exponential fitting to the tails of  density distribution
with $\sim \exp(-|x|/L)$ } \label{fig4}
\end{figure}

Anderson localization in a weakly disordered potential is
characterized by a long  exponential tail of the localized
state. For  collisionally homogeneous condensates,
the experimental \cite{roati} and theoretical \cite{adhikari1}
investigations have demonstrated that the localized BECs have an
exponential tail for weakly-interacting or non-interacting BEC in a
quasi-periodic OL. In order to observe the effect of spatially
modulated nonlinearity on the tail region, we plot in Fig.
\ref{fig4} (a) the  density distribution $|u|^2$
of the stationary BEC on log scale. As we see, the long exponential
tail extends from $x\approx\pm2$ to  $x\approx\pm20$,
whereas the central part of 
density for $|x|<2$
represents  a Gaussian distribution.
By an exponential fitting  of the exponential function $\sim \exp(-|x|/L)$
to the tails of  density
distribution, the localization length $L$
versus $\varepsilon$ is illustrated  in Fig. \ref{fig4} (b) which  shows
that
$L$ increases nonlinearly with $\varepsilon$. An increase in $\varepsilon$
represents a decrease in disorder thus  resulting in larger values of
localization length.

\section{Dynamics of Localized State}
\label{IIII}

To get further insight into the effects of the spatially modulated
nonlinearity $\varepsilon$
on the localized states, we now study  some
dynamics of the localized state. First, we study
numerically the oscillation  of the localized states in an external
potential.
According to Eqs. (\ref{veff}) - (\ref{veffV}), the motion of the
localized BEC can be approximately regarded as that of a
particle inside an effective potential $V_{\rm eff}$. Because
of the exponential tail and elasticity of the localized state,
although the variational results may not be good for the
dynamical evolution, they can provide a qualitative  physical
understanding of the dynamics using the effective potential.
To study the motion of the
localized state,  first we create  a stationary  localized BEC
with spatially modulated nonlinearity in the
bichromatic OL. Successively, at $t=0$, we suddenly introduce an
initial {momentum  $p_0=0.1$ by $u(x)\to u_0(x)\exp(ip_0x)$,}
where $u_0$ is the
wave function of the stationary  localized BEC. From the
experimental point of view, the initial momentum  can be given by
suddenly moving the OL \cite{moving}.  It is found that after the perturbation,
the  density
envelope suffers an abrupt change but remains localized.
Actually, the localized BEC is an elastic object and not a rigid
one. Hence,  both the center and the density distribution 
of the localized state
perform oscillations.

\begin{figure}
\begin{center}
\includegraphics[width=\linewidth]{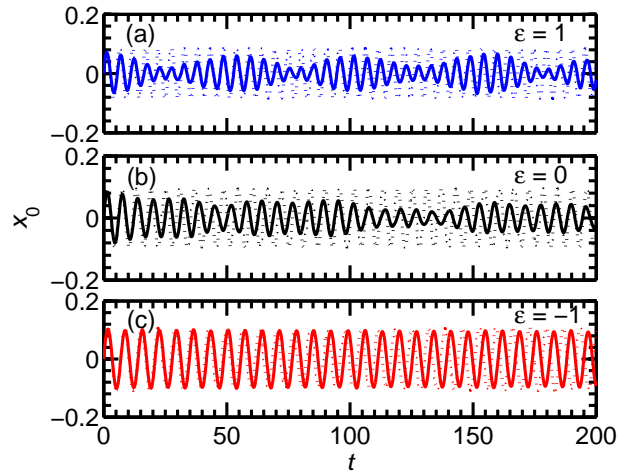}
\end{center}

\caption{(Color online) The center of the 
localized state $x_0$ versus time $t$ (both dimensionless)
during the location oscillation
initiated by suddenly introducing an initial momentum 
$p_0=0.1$ by the transformation $u\to u_0\exp (ip_0x)$ 
for 
$\varepsilon = $ (a) 1,
(b) 
0, and  (c) $-1$: numerical result (full line) and 
variational result (dotted line).  } \label{fig5}
\end{figure}

The evolution of the  center
 $x_0$ of the localized BEC as obtained from numerical  
simulation  (full line)
is shown in Fig. \ref{fig5}
where $x_0$ is obtained
by instant Gaussian-function fitting to the central region of the
 density distribution. The variational results are also shown in this figure.  
The top, middle and bottom panels
correspond to $\varepsilon=1, 0$, and $-1$, respectively. We find
that, in general,  the oscillation of the localized state could be
quasi periodic
after an initial damping, which  can be explained on the basis of energy
conservation. Because of the deformation of the  density
envelope, a part of the initial kinetic energy is converted into
elastic energy of  deformation
and this  leads to the damped oscillation. 
{{This deformation will be larger for large $\varepsilon (=1)$,
when the localized state is loosely bound with large exponential tail 
(viz. Fig. \ref{fig4} (a)). Consequently, an oscillation of the center 
$x_0$ with rapidly varying amplitude is found denoting easy periodic 
transfer of energy between location oscillation and deformation. 
For small $\varepsilon (=-1)$, the localized state 
is more compact and tightly bound,  so that it can be treated like a rigid 
object. The exchange of energy is less probable in that case, and a periodic 
oscillation of the center $x_0$ with constant amplitude is found.}} 
Eventually, the energy of deformation 
is liberated leading to an increase in the amplitude of oscillation.
During the subsequent
oscillation cycle, the conversion between the kinetic
energy and  elastic strain energy  causes the
quasi-periodical movement of the localized state.
In the  variational formulation the exchange of kinetic energy to the 
energy of deformation is not allowed and the resultant oscillation is of 
a fixed amplitude without damping. Nevertheless, the numerical frequency
of location oscillation  
is in agreement with the variational frequency within an estimated error 
of about 2.5\%. 
As pointed out in Sec. \ref{III}, a positive
$\varepsilon$ leads to a tighter trapping and vice versa.
Then, with the same initial velocity, the tighter trapping causes
the localized state to oscillate with larger frequency  and smaller amplitude,
and a weaker trapping leads to  a smaller frequency  and larger amplitude of
oscillation as indicated in the numerical results of Fig. \ref{fig5}. The 
variational frequency follows the same trend as $\varepsilon$ is changed 
from positive values to negative values. 


\begin{figure}
\begin{center}
\includegraphics[width=\linewidth]{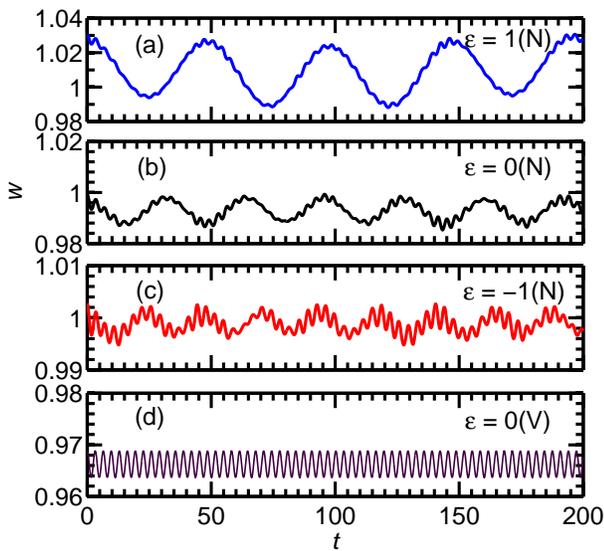}
\end{center}

\caption{(Color online) Numerical (N) 
result for  dimensionless pulse width $w$ of the localized state
versus  dimensionless 
time $t$ during breathing oscillation initiated by suddenly changing 
the strength of the secondary lattice $s_2$ from 3 to 3.5 
for  $\varepsilon = $  (a) $1$ ,
(b) 0 , and (c) $-1$.  We also show the variational result (V)  of pulse width
$w$   
from a solution of Eqs. (\ref{width}) and (\ref{chirp}) with 
condition $w(t=0)= 0.9688$ and $\beta(t=0)=0$.} \label{fig6}
\end{figure}

Next we consider a
breathing oscillation of the localized BEC,
started  by suddenly changing the strength  of the secondary
lattice $s_2$ from 3 to 3.5 at $t=0$.
We investigate how the breathing oscillation of the localized
BEC is changed by the spatially modulated nonlinearity.
Then, the nonlinearity in Eq.
(\ref{eq1}), $g(x)$, also changes  with the new OL potential. We
present numerical results in Fig. \ref{fig6} (a), (b), and (c)
for the time evolution
of the width  $w$ of the localized state for $\varepsilon
=1,0$ and $-1$, respectively. 
The variational equations (\ref{width}) 
and (\ref{chirp}) were  solved to obtain the oscillation of the 
central part and this result is shown in Fig. \ref{fig6} (d)
for $\varepsilon=0$. (The Gaussian variational ansatz without 
any exponential tail
only carries 
information about the central part.) 
There are two regions of the localized state which oscilate with two 
distinct frequencies: (i) the central region with Gaussian distribution 
and (ii) the outer tail with exponential distribution. The net 
result is the 
harmonic oscillation of the width with 
a small frequency and large amplitude due 
to the oscillation of the 
exponential  tail, which is  modulated by rapid oscillation coming from the 
central Gaussian part. 
The variational frequency as obtained from Fig. \ref{fig6} (d) is 
found to be identical with the frequency of modulation of the 
numerical width, which confirms that the modulation in Figs.
 \ref{fig6} (a), (b) and (c) is coming from the oscillation of the 
central part of the condensate.
The renewed oscillation in Figs. \ref{fig5}  and \ref{fig6} confirms the
stability of the
stationary localized BECs.

\section{SUMMARY}
\label{IIIII}

In this paper, using the numerical and variational solution of the
time-dependent GP equation, we studied the stationary and dynamic
characteristics of a cigar-shaped localized BEC with spatially
inhomogeneous nonlinearity in a bichromatic quasi-periodic 1D OL
potential. This investigation reveals that the spatially inhomogeneous
nonlinearity produces a pseudo-potential which changes the strength
of the disorder and the height of the quasi periodic effective
potential felt by the localized BEC. With a larger spatially
modulated coefficient $\varepsilon$,
the localization length and Hamiltonian will be larger.

 We also study the stability of the stationary localized
state using the linear stability analysis and find it is dynamically
stable under small perturbations. The stability is also verified by
numerical simulation. In respect to  dynamics, we investigate
the location oscillation (oscillation of the center)
and breathing oscillation of
the localized BEC, and find that both  oscillations are quasi
periodic because of the quasi periodic effective potential.
The frequency of quasi periodic oscillations of the center of the BEC
increases as
 $\varepsilon$ increases.   For the breathing
oscillations, the two exponential
tails also are symmetric around the center at
$x=0$. The present study is useful for an  understanding of the
statics and dynamics  of
Anderson localization and for planing new experiments with collisionally
inhomogeneous BEC.

\acknowledgments

FAPESP (Brazil) and CNPq (Brazil)
provided partial support.



\end{document}